\documentclass[aps,pre,twocolumn]{revtex4}
\usepackage{graphicx} 
\usepackage{tikz-cd}
\usepackage{amsmath}
\usepackage{amsfonts}
\usepackage{amssymb}
\usepackage{adjustbox}
\usepackage{appendix}
\usepackage{xcolor}

\begin{document}
\title{Defect Dynamics in Cholesterics: Beyond the Peach--Koehler Force}
\author{Joseph Pollard}
\email{joe.pollard@unsw.edu.au}
\affiliation{School of Physics, UNSW, Sydney, NSW 2052, Australia.}
\affiliation{EMBL Australia Node in Single Molecule Science, School of Medical Sciences, UNSW, Sydney, NSW 2052, Australia.}
\author{Richard G. Morris}
\email{r.g.morris@unsw.edu.au}
\affiliation{School of Physics, UNSW, Sydney, NSW 2052, Australia.}
\affiliation{EMBL Australia Node in Single Molecule Science, School of Medical Sciences, UNSW, Sydney, NSW 2052, Australia.}
\affiliation{ARC Centre of Excellence for the Mathematical Analysis of Cellular Systems, UNSW Node, Sydney, NSW 2052, Australia.}

\begin{abstract}
The Peach--Koehler force between disclination lines was originally formulated in the study of crystalline solids, and has since been adopted to provide a notion of interactions between disclination lines in nematic liquid crystals. Here, we argue that the standard formulation of this interaction force seemingly fails for materials where there is a symmetry-broken ground state, and suggest that this is due to the interaction between disclination lines and merons: non-singular yet non-trivial topological solitons. We examine this in the context of chiral nematic (cholesteric) liquid crystals, which provide a natural setting for studying these interactions due to their energetic preference for meron tubes in the form of double-twist cylinders. Through a combination of theory and simulation we demonstrate that, for sufficiently strong chirality, defects of $+1/2$ winding will change their winding through the emission of a meron line, and that interactions between the merons and defects dominate over defect-defect interactions. Instead of Peach--Koehler framework, we employ a method based on contact topology---the Gray stability theorem---to directly calculate the velocity field of the material. We apply our framework to point defects as well as disclination lines. Our results have implications not just for chiral materials, but also for other phases with modulated ground states, such as the twist-bend and splay-bend nematics. 
\end{abstract}

\maketitle

\section{Introduction}
Disclination lines are the fundamental topological defects of nematic liquid crystals in three-dimensional geometries. As centres of strong elastic distortion they are the dominant factor in the material's dynamics, and as such have been studied extensively~\cite{kleman_classification_1977, kleman_relationship_1977, mermin_topological_1979, volovik_topological_1983, kleman_points_1983, janich_topological_1987, kleman_defects_1989, kleman_disclinations_2008, copar_topology_2014, long_geometry_2021, alexander_entanglements_2022, schimming_kinematics_2023, long_applications_2024}. Amongst other things, they govern the isotropic-nematic transition and subsequent coarsening processes~\cite{kosterlitz_ordering_1973, renn_abrikosov_1988, chuang_coarsening_1993}, as well as determine the behaviour of stable nematic textures, including their response to elastic stresses and external fields~\cite{taylor_mechanism_1934, poulin_novel_1997, skarabot_two-dimensional_2007}.  

In crystalline materials, the standard approach to studying interactions between defects involves the Peach--Koehler force, an interaction force between disclination lines which is analogous to the Lorentz force in electromagnetism~\cite{peach_forces_1950}. It has since been adopted to describe the interactions between disclination lines in nematic liquid crystals~\cite{eshelby_force_1980, kleman_points_1983}. To apply the Peach--Koehler framework, we fix a set of defects and minimise the free energy subject to the constraint of the material having these defects. The resulting stresses and the Burgers vector of the disclination line are then used to compute the forces experienced by the disclination lines.
It is also important to include additional forces, such as a line tension or terms coming from boundary conditions~\cite{long_applications_2024}.
Notably, this framework has been applied to study disclination motion in nematic materials in the one elastic constant approximation; a setting in which it has been very successful, yielding predictions that agree with simulation and experiment~\cite{long_geometry_2021, schimming_kinematics_2023, long_frank-read_2024, long_applications_2024}.

However, to our knowledge, the predictions of this theory have not not been studied in depth in more complex settings, for example in materials where the single elastic constant approximation is strongly violated---for example, in lyotropic chromonic materials the elastic anisotropy lies at the heart of the interesting phenomenology, causing them to form chiral structures even though the molecules are not themselves chiral---or materials with broken symmetry in the ground state, such as cholesterics (chiral nematics)~\cite{beller_geometry_2014} and twist-bend and splay-bend materials~\cite{jakli_physics_2018}.
In these cases, there are three broad challenges to applying the Peach--Koehler theory. First, at a practical level, there is difficulty in computing the minimisers of the energy, which is relatively straightforward in the one elastic constant approximation but not when there is elastic anisotropy or broken-symmetry terms in the energy. Second, and more fundamentally, non-uniform global structures selected for by the broken symmetry terms can also cause issues---for example, cholesterics select for helical winding in the director field, including along disclination lines, and this is not captured by the Peach--Koehler force. Finally, and most pertinent to this paper, these classes of material exhibit stable topological solitons, called merons and Skyrmions. In a recent paper, we described the topological and geometric aspects of defect-defect interactions, and showed that merons play a fundamental role in mediating interactions~\cite{pollard_morse_2024}. Defect-meron interactions are not captured by the Peach--Koehler force, at least not in its standard formulation. Therefore, to go beyond the results of our previous work~\cite{pollard_morse_2024}---which focusses on topological and structural properties which are agnostic to the actual physics of these materials---we must understand the role that elastic forces play in these interactions. 

The purpose of this paper is to begin an investigation of defect-defect and defect-meron interactions in materials that cannot be a approximated with one elastic constant. We choose to examine these issues in a particular class of materials, the chiral nematic liquid crystals, also called cholesterics. There are two key reasons why these are a convenient class of materials in which to study this problem. Firstly, chirality stabilises meron tubes in the form of `double-twist cylinders', which are energetically favoured over a meron-free texture for sufficiently strong chirality~\cite{wright_crystalline_1989}---thus, the issue of how merons interact with defects and mediate interactions between defects is of genuine practical importance in cholesterics. Secondly, the special topology of the cholesteric phase allows for their study via the machinery of contact topology~\cite{geiges_introduction_2008, machon_contact_2017, pollard_point_2019, eun_layering_2021, han_uniaxial_2022, pollard_contact_2023, pollard_escape_2024}, making them analytically tractable in a way that other classes of material are not. 

The remainder of this paper is organised as follows. We review the topology of cholesterics in Section~\ref{sec:cholesterics}. By comparing the predictions of the Peach--Koehler theory with numerical simulations, we show in Section~\ref{sec:disclination_motion} that this theory appears not to give good predictions for the motion and behaviour of defects in a chiral material when the chirality is sufficiently strong; we see substantially different behaviour from an achiral system. Chiral disclinations with an initially $+1/2$-winding profile change their winding by the emission of a meron line (double-twist cylinder), and that a winding of the director profile along the disclination may cause it to buckle into a helix. This last case cannot be understood within the Peach--Koehler framework, as the Peach--Koehler force is entirely insensitive to profile winding~\cite{long_geometry_2021}. These observations suggest the Peach--Koehler interaction force between disclinations can be overcome by a force arising from the merons. Interestingly, merons were originally introduced in quantum chromodynamics to explain the phenomenon of quark confinement~\cite{callan_toward_1978}---a similar effect occurs in cholesterics, with merons acting to screen interactions between disclination lines. 

Instead of using the Peach--Koehler framework to analyse these materials, in Section~\ref{sec:gray_stability} we adopt an approach based on contact topology. Under the assumption that the material maintains a consistent sense of handedness, the Gray stability theorem~\cite{gray_global_1959, geiges_introduction_2008} allows us to explicitly calculate the velocity field of the material. This technique has previously been applied to defect-free cholesterics~\cite{machon_contact_2017}, but we apply it to the case where the material contains disclination lines for the first time. We show that the predictions of this theory correspond much more closely to what is seen in simulation and experiment. Flow-field calculations show that meron tubes are driven to expand, and it is this which drives the motion of defects in a cholesteric, not Peach--Koehler-type interactions between the defects themselves. This behaviour appears generic in a cholesteric, and the change in winding of $+1/2$-lines appears to be the fundamental mechanism underlying the formation of double-twist regions from an initially random state. 

One advantage to an approach based on contact topology is that it doesn't just allow for an analysis of disclination lines, but can also be applied to geometric structures such as cholesteric `layers', topological solitons, and even point defects. In Section~\ref{sec:hedgehog} we analyse the motion of a radial hedgehog defect in a droplet, and again show that our theory corresponds closely to what is observed in simulation and experiment--namely, that a radial hedgehog will displace from the centre of the droplet towards the boundary. In this particular example the material does not (and cannot~\cite{pollard_point_2019}) maintain a consistent sense of handedness throughout the droplet or even in a neighbourhood of the defect. The fact that our technique still applies when this constraint is relaxed suggests that we may well be able to apply this framework to a broader class of materials which are of current experimental interest, for example lyotropic chromonics, as well as the twist-bend and splay-bend nematics. We close in Section~\ref{sec:discussion} with a discussion of our results and the possibilities for future work.

\section{Background on Cholesterics and the Peach--Koehler Force}
\label{sec:cholesterics}
Chiral materials furnish an especially rich source of topological and geometric phenomena, and liquid crystal materials comprised of chiral molecules (cholesterics) are no exception. In this section we review both the topology and the elasticity of these materials, and explain how the deceptively simple chiral symmetry-breaking leads to substantially different behaviour from an achiral material. 

The free energy for a cholesteric liquid crystal in the one elastic constant approximation is,
\begin{equation} \label{eq:frank_energy}
    E_\text{Frank} = \frac{K}{2}\int |\nabla {\bf n}|^2 + 2q_0 {\bf n} \cdot \nabla \times {\bf n} + q_0^2.
\end{equation}
Here, $K$ is the elastic constant and $q_0$ is a symmetry-breaking parameter with dimensions of inverse length which sets the pitch length $p = 2\pi/q_0$, the distance over which the director rotates by $2\pi$. The sign of $q_0$ determines the sense of this rotation. We take $q_0 > 0$ for a right-handed material, but emphasise that all of our results are equally valid for a left-handed material. We recover the energy for an achiral system (in the one constant approximation) by setting $q_0=0$. 

This energy favours nonzero twisting of the material, ${\bf n} \cdot \nabla \times {\bf n} \sim - q_0$, with molecules rotating along a given direction ${\bf p}$ called the pitch axis~\cite{beller_geometry_2014, machon_umbilic_2016}. Making the assumption that the material is everywhere chiral, ${\bf n} \cdot \nabla \times {\bf n} \neq 0$, allows for an analysis of cholesterics using the machinery of contact topology~\cite{machon_contact_2017, pollard_point_2019, eun_layering_2021, han_uniaxial_2022, pollard_contact_2023, pollard_escape_2024}, with results that compare well with simulation and experiment. In this work, we further develop the ideas of Machon~\cite{machon_contact_2017} to understand how chirality influences structural changes to defects.

We remark that~\eqref{eq:frank_energy} is also the energy for a ferromagnet with the Dzyaloshinskii--Moriya interaction interaction, with ${\bf n}$ now a magnetic polarisation field. Since the magnetic field is orientable disclination lines do not arise in these materials, but merons, Skyrmions, and point defects (`Bloch points') do. Thus, some parts of our results will apply to these materials as well. There is a further potential application to superfluid phases of helium-3, which can form chiral structures and exhibit point defects, merons/Skyrmions, and string-like objects analogous to disclination lines~\cite{volovik_string_2020, volovik_composite_2020}.

\subsection{The Contact Topology of Cholesterics and Their Disclinations}
We briefly review the contact topology of chiral materials. The first key distinction between chiral and achiral materials is that the former have a binary topological invariant called `overtwistedness'. A cholesteric texture is overtwisted if it contains a `double-twist cylinder', the core of which is a meron with $+1$ winding and a rotational structure (a `Bloch' meron), and is tight if no such structures exist in the material. The standard cholesteric helix, 
\begin{equation} \label{eq:standard}
    {\bf n}_q = \cos qz \, {\bf e}_x + \sin qz {\bf e}_y,
\end{equation}
is tight. Meron/Skyrmion lattices and blue phases~\cite{wright_crystalline_1989} are examples of overtwisted textures. For sufficiently high chirality overtwistedness is energetically preferred, as double-twist regions have lower energy than the director~\eqref{eq:standard}.

Correspondingly, disclination lines in cholesterics can be broken down into two topologically-distinct classes depending on whether a local neighbourhood is tight or overtwisted~\cite{pollard_contact_2023}. The tight disclinations are exactly those that possess a well-defined winding number along the entire length of the line---that is, on any cross-sectional disk we can consider the winding of the director around the boundary of the disk, and for a tight disclination this number is constant and independent of the choice of disk---and they arise in the majority of well-known cholesteric textures featuring defects, including meron lattices and blue phases; the overtwisted disclinations have changes in winding, which may or may not be mediated by meron tethers attached to the line~\cite{pollard_contact_2023, pollard_escape_2024, pollard_morse_2024}, and will result from crossing tight disclination lines~\cite{poenaru_crossing_1977}. 

Examples of overtwisted disclinations with attached meron tethers have been realised in recent experiments~\cite{wu_topological_2024}, however we will only discuss the tight class of disclination lines here, as these are the stable defects occuring in the majority of cholesteric textures. These further split into the $\tau$-lines, which are also singularities of the cholesteric pitch axis, and the $\chi$-lines, along which the pitch axis is nonsingular~\cite{beller_geometry_2014, pollard_contact_2023}. The former are analogous to edge dislocations in a crystal or in a smectic---as they are singular for the pitch axis, they are also defects for the cholesteric's layer structure---while the latter possess a helical winding and are analogous to screw dislocations in a smectic. 

A tight director field that is constrained to be free from defects cannot be made overtwisted without the creation of regions of reversed handedness, and vice-versa. Thus, in defect-free textures tightness is a condition which is protected by a chiral energy barrier roughly proportional to $Kq_0/2$ times the size of the region in which the constraint ${\bf n} \cdot \nabla \times {\bf n} \neq 0$ is violated~\cite{machon_contact_2017}. Only in regions of sufficiently large bend or splay distortion can this occur. The $\chi$ and $\tau$ classes of disclination are also topologically distinct, and for the $\chi$ class the helical winding $q$ is invariant under homotopies which preserve the condition ${\bf n} \cdot \nabla \times {\bf n} \neq 0$ and create no additional defects---this winding number is therefore also protected by a chiral energy barrier~\cite{pollard_contact_2023}.

Defects open up an additional pathway by which a tight material can become overtwisted. It is possible for a tight $\pm 1/2$ disclination (of either $\chi$-type or $\tau$-type) to convert into a $\mp 1/2$ disclination line by the emission of a $\pm 1$-winding meron line~\cite{pollard_morse_2024}. Meron lines of $-1$-winding and those of $+1$-winding with a radial profile (`Neel' merons) are energetically costly, and thus we do not expect them to spontaneously nucleate. However, it is well-known that meron lines with a rotational profile---double-twist cylinders/Bloch merons---are energetically favoured in a cholesteric~\cite{wright_crystalline_1989}, and so disclination lines of $+1/2$-winding can serve as sources for the nucleation of topological solitons within a cholesteric material via their conversion into $-1/2$ disclinations---we discuss this further in later sections. It is also possible for a pair of $+1/2$ disclinations to merge to create a $+1$ singular line which then `escapes' to form a meron tube, provided a force acts on them which overcomes the natural repulsion of like-sign defects. This merging can occur along the entire length of the disclinations, removing them entirely, but also potentially along just a segment of the disclinations, resulting in a pair of disclinations with a meron tether~\cite{pollard_escape_2024, pollard_morse_2024, wu_topological_2024}. 

Since the presence of these double-twist cylinders is exactly the condition of overtwistedness, we see that defects serve as mediators for the transition between tight and overtwisted textures. Double-twist cylinders are the energetically-preferred structure in a cholesteric~\cite{wright_crystalline_1989}, so understanding the defect- and meron-mediated transition from a (tight) helical state to an (overtwisted) state with a lattice of double-twist cylinders is of some interest beyond the question of disclination motion.

\subsection{Cholesteric Elasticity}
To derive elastic forces such as the Peach--Koehler force, we require appropriate definitions of the stress and strain in a cholesteric. First, we briefly review Lagrangian elasticity in a tensor-based framework, following Eshelby~\cite{eshelby_elastic_1975, eshelby_force_1980}. Suppose the material can be described by a displacement field ${\bf u} = u_i {\bf e}_i$ written in a coordinate frame and also assume that there exists a stored energy function (energy density) $W(u_i, \partial_j u_i)$ depending on ${\bf u}$ and its gradient. We define the material strain to be $\epsilon_{ij} = \partial_j u_i$. The Euler--Lagrange equation governing the evolution of the displacement field. 
\begin{equation}
    \partial_j \frac{\partial W}{\partial \epsilon_{ij}} - \frac{\partial W}{\partial u_i} = \frac{\partial u_i}{\partial t},
\end{equation}
with summation over repeated indices implied. Write 
\begin{equation} \label{eq:stress_definition}
    \sigma_{ij} = \frac{\partial W}{\partial \epsilon_{ij}}, \ \ \ \ \ f_i = \frac{\partial W}{\partial u_i}.
\end{equation}
Then $\sigma_{ij}$ is the first Piola--Kirchoff stress tensor, and its divergence gives rise to a force associated with with momentum conservation, while ${\bf f} = f_i {\bf e}_i$ is a force which does not necessarily arise as the divergence of a stress-tensor. The Euler--Lagrange equation for a steady state with $\partial_t {\bf u} = 0$ is then the force-balance condition $\nabla \cdot \sigma = {\bf f}$, which characterises the equilibrium configurations of the material. 

We may apply this framework to a nematic by taking the director ${\bf n} = n_i {\bf e}_i$ to stand for the displacement field and choosing our energy functional to be the Frank energy. This suggests a choice of strain $\epsilon_{ij} = \partial_j n_i$. We may define the stress $\sigma$ and the force ${\bf f}$ as in \eqref{eq:stress_definition}. In components, the Frank energy density is
\begin{equation} \label{eq:energy_density}
    \begin{aligned}
        W &= \frac{K}{2}\left((\partial_j n_i)^2 + 2q_0\gamma_{kji}n_k \partial_j n_i \right), \\
        &= \frac{K}{2}\left(\epsilon_{ij}^2 + 2q_0\gamma_{kji}n_k\epsilon_{ij} \right),
    \end{aligned}
\end{equation}
where $\gamma_{ijk}$ is the Levi-Civita tensor. A short calculation then gives
\begin{equation} \label{eq:stress_force_chiral}
    \sigma = K(\nabla {\bf n} + q_0 J), \ \ \ \ \ {\bf f}_\text{ch} = Kq_0 \nabla \times {\bf n},
\end{equation}
where $J = {\bf n}\times$ is anticlockwise rotation about the director, which can be straightfowardly viewed as a tensor with components $J_{ij} = \gamma_{kji}n_k$. A simple calculation gives $\nabla \cdot J = -\nabla \times {\bf n}$, and so the force balance $\nabla \cdot \sigma - {\bf f}_\text{ch} = 0$ results in the familiar Euler--Lagrange equation governing the dynamics of energy-minimisation in the cholesteric phase~\cite{degennes_physics_2013}, 
\begin{equation} \label{eq:EL}
    \partial_t {\bf n} = K\left(\nabla^2 {\bf n} -2q_0\nabla \times {\bf n}\right).  
\end{equation}
There is a subtlety however: in a nematic material we should not identify the strain with $\nabla {\bf n}$, as this is not invariant under the nematic symmetry ${\bf n} \mapsto -{\bf n}$. We can remedy this by instead considering $\epsilon = J\nabla {\bf n}$. By differentiating the Frank energy density~\eqref{eq:energy_density} with respect to this choice of strain, we find that the correct form of the stress is in fact
\begin{equation} \label{eq:stress_chiral}
    \sigma = K(J\nabla {\bf n} - q_0(I - {\bf n} {\bf n}) ), 
\end{equation}
where $I$ denotes the 3D identity tensor. The tensor $(I - {\bf n} {\bf n})$ is simply the projection operator onto the planes orthogonal to ${\bf n}$, with components $\delta_{ij} - n_i n_j$, where $\delta_{ij}$ is the Kronecker delta. This is obtained by applying $J$ to the stress tensor in~\eqref{eq:stress_force_chiral}. We should also consider the `torque' $J{\bf f}_\text{ch}$ rather than the force itself, as the latter is not invariant under the nematic symmetry. 

An alternative approach to cholesteric elasticity involves the introduction of a `cholesteric derivative', a connection $D$ which is still compatible with the Euclidean metric but has torsion. This connection is defined by,
\begin{equation} \label{eq:cholesteric_connection}
    (D_i {\bf n})_j = \nabla_i n_j + q_0\gamma_{ijk}n_k,
\end{equation}
or equivalently $D {\bf n} = \nabla {\bf n} -q_0 J$. We can then consider the energy density 
\begin{equation} \label{eq:cholesteric_energy}
    W^\prime = \frac{K}{2}|D{\bf n}|^2,
\end{equation}
which `hides' the broken symmetry inside the derivative. The free energy $W^\prime$ is minimised by a director satisfying $D{\bf n} = 0$. This constraint, which selects for double-twist---i.e., for overtwisted director fields---cannot be satisfied over an extended region of flat space~\cite{sethna_relieving_1983, wright_crystalline_1989}, a form of geometric frustration which can only be relieved by the introduction of defects. 

The energy density $W^\prime$ differs from the Frank energy density $W$ of Eq.~\eqref{eq:energy_density} by the saddle-splay term. This saddle-splay is a total divergence which does not contribute to the Euler--Lagrange equation in the bulk, but nonetheless plays an important role in the elasticity of cholesterics, especially their defects~\cite{sethna_relieving_1983, wright_crystalline_1989, machon_umbilic_2016, selinger_interpretation_2018}. In terms of this energy density, the strain is $\epsilon = J D{\bf n}$, and the associated stress tensor is $\sigma = KJD{\bf n}$---this then yields the same stress tensor as we get from the Frank energy density, Eq.~\ref{eq:stress_chiral}.

\subsection{The Ericksen and Peach--Koehler Forces}
Computing the Peach--Koehler interaction force between defects in a nematic requires us to know the Burgers vector of the defect. Regardless of whether we are dealing with $\tau$- or $\chi$-lines, liquid crystal defects are analogous to screw dislocations in a crystal. For a defect in a liquid crystal with winding $k$ the Burgers' vector is therefore ${\bf B} = 2\pi k \Omega$, where $\Omega$ is the `rotation vector', defined so that the director field lies everywhere in the planes orthogonal to $\Omega$ along the disclination~\cite{friedel_buckling_1969, kleman_points_1983, long_geometry_2021}. 

In a cholesteric, this definition leaves something to be desired. The Burgers vector can distinguish between tight and overtwisted disclinations, as the latter class must have a change in winding, which in turn corresponds to points where $\Omega$ is orthogonal to the disclination (which we may call `Legendrian points', after a related concept in contact topology~\cite{geiges_introduction_2008}). However, all tight disclinations have a neighbourhood in which the director can be homotoped so that the rotation vector is $\Omega = \pm {\bf t}$, for ${\bf t}$ the tangent vector to the line~\cite{pollard_contact_2023}. Thus, the Burgers vector does not distinguish between edge-like ($\tau$) and screw-like ($\chi$) disclinations, despite their behaviour being quantitatively different, and nor does it `see' the topological invariant $q$. The insensitivity of the Burgers vector to helical windings is a known issue~\cite{long_geometry_2021}. It may be possible to develop an alternative approach to the Burgers vector which accurately captures this distinction in a cholesteric. This approach, outlined briefly in our previous work~\cite{pollard_morse_2024}, is based on the method of convex surface theory for contact topology~\cite{geiges_introduction_2008} and conceptually much closer to the Volterra process construction of defects~\cite{kleman_points_1983}. Studying possible generalisations or adaptions of the Burgers vector concept is beyond the scope of this work, and we continue to use the standard definition in the remainder of this article. 

If ${\bf t}$ denotes the tangent vector to the line, then the Peach--Koehler force (per unit length) is~\cite{long_geometry_2021}
\begin{equation} \label{eq:peach-koehler}
    {\bf f}_\text{PK} = ({\bf B} \cdot \sigma)\times {\bf t}.
\end{equation}
Evaluating this along a given disclination line gives the instantaneous velocity of that defect at each point along the line. Integrating over a tube enclosing the dislocation gives the overall force acting on it. 

An alternative approach to deriving this force was given by Eshelby~\cite{eshelby_elastic_1975, eshelby_force_1980}. The stress and strain can be packaged into the Ericksen stress tensor $\rho$, which has components, 
\begin{equation} \label{eq:Ericksen_stress}
    \rho_{ij} = (W-p_0)\delta_{ij} -\epsilon_{ki}\sigma_{kj},
\end{equation}
where $p_0$ is a uniform hydrostatic pressure. This is equivalent to the energy momentum tensor in a nematic~\cite{eshelby_force_1980}, differing only by the pressure term. This tensor can be integrated over a surface $S$ to give the `Ericksen force' exerted on the material inside $S$ by the medium outside it. Let ${\bf N}$ be the normal to $S$ and $dA$ its area element. The Ericksen force ${\bf f}_E$ is given by 
\begin{equation} \label{eq:Ericksen_force}
    {\bf f}_E = \int_S \rho \cdot {\bf N} dA.
\end{equation}
We can consider the force density $\rho \cdot {\bf N}$ restricted to the disclination to be a force per unit length, and it is equal to the Peach--Koehler force (per unit length) of Eq.~\eqref{eq:peach-koehler}~\cite{eshelby_force_1980}. 

To obtain the force on a disclination line, we must take $S$ to be a tube enclosing the disclination. This approach makes explicit a fundamental assumption that underlies the entire theory: for the force to be well-defined, the integral must be independent of the choice of this surface. This will be the case exactly when $\rho$ is divergence-free, and this in turn occurs only when the director minimises the energy with density $W$, and so this assumption is necessary for the force to be well-defined. It is well known that the Peach--Koehler theory gives incorrect predictions of disclination line motion when this assumption is not met. For example, if a pair of disclination lines of opposite winding have profiles that are `twisted' relative to one another, not minimising the free energy, then the defects move towards one another on curved trajectories, while the Peach--Koehler theory predicts that they will move along the straight line segment between them~\cite{tang_orientation_2017, schimming_kinematics_2023}. 

In an achiral nematic in the one elastic constant approximation this assumption means finding a director field which is harmonic, $\nabla^2 {\bf n} =0$, which is not generally difficult. However, if we wish to generalise the theory to treat chiral materials or materials with distinct elastic constants then finding an energy minimising director analytically is now more of a challenge. This limits practical application of the theory beyond the assumption of ordinary nematics in the one elastic constant approximation. Alternative approaches which do not rely on this assumption (and which also apply to point defects) have been discussed in the literature~\cite{schimming_kinematics_2023}, but these can also become very complicated.

\section{Disclination Motion in a Cholesteric}
\label{sec:disclination_motion}

\begin{figure*}[t]
\centering
\includegraphics[width=0.98\textwidth]{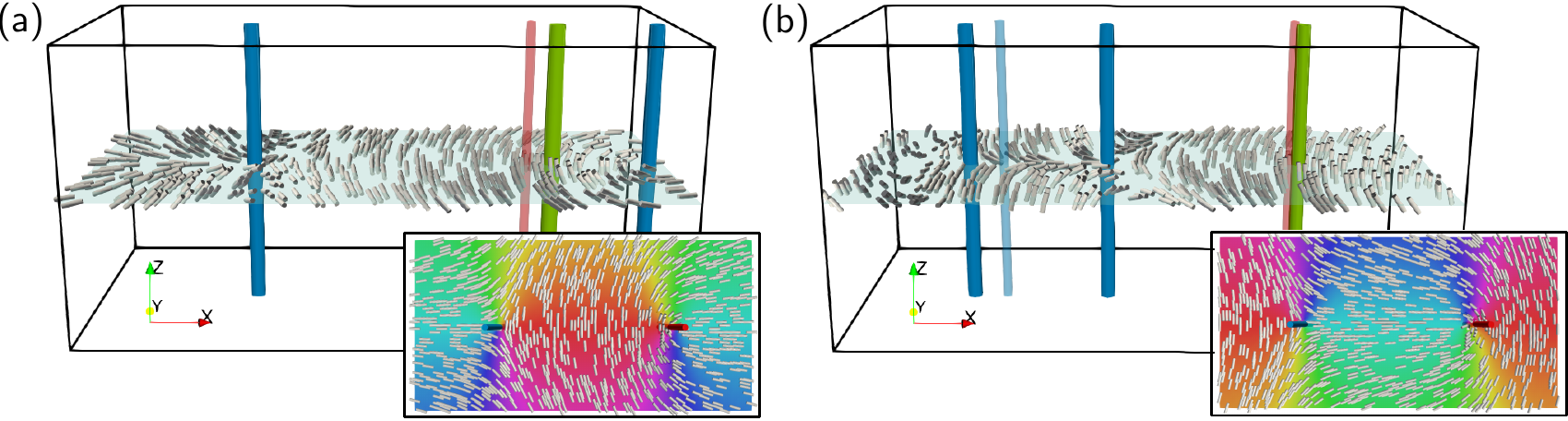}
\caption{The motion of $\tau$-lines in a cholesteric depends on the winding and orientation. We consider a pair of disclination lines, one with winding $-1/2$ (blue) and one with winding $+1/2$ (red). Lines with winding $+1/2$ move along the direction of their `tail', and rapidly change their winding to $-1/2$ via the emission of a $+1$-winding rotation meron line (green), which has the structure of a double-twist cylinder. The motion of the $+1/2$-winding line can therefore be either (a) away from, or (b) towards the $-1/2$-winding line, depending on the relative orientation of the profiles. This is in contrast to the achiral nematic case, where opposite winding disclinations move together and annihilate. We show the initial positions of the lines as transparent tubes, and the final equilibrium positions as solid tubes. In each case we show an inset with the initial director across a surface, coloured according to the angle between the $x$ and $y$ components. The simulations shown are performed when the pitch $p$ is equal to the height of the simulation box.}
\label{fig1}
\end{figure*}

We now examine interactions between disclination lines in a chiral material through a combination of the Peach--Koehler theory and simulation. The simplest nontrivial case for us to study consists of a pair of straight, parallel disclination strands along lines $L_1, L_2$ with winding numbers $k_1, k_2$, respectively. It can be shown that a director field minimising the energy for this configuration in the achiral case ($q_0=0$) is ${\bf n} = \cos \phi \, {\bf e}_x + \sin \phi \, {\bf e}_y$, with $\phi$ the harmonic function~\cite{tang_orientation_2017, long_geometry_2021},
\begin{equation} \label{eq:two_disclinations}
    \phi = qz + k_1 \tan^{-1}\left(\frac{y}{x+s} \right) + k_2 \tan^{-1}\left(\frac{y}{x-s}\right),
\end{equation}
Here, $L_1$ is at the position $(-s, 0)$ and $L_2$ at $(s,0)$. The parameter $q$ allows for a helical rotation of the disclination profiles along the $z$-axis. We can easily calculate the Peach--Koehler force exerted on $L_2$ in an achiral system, see for example Long \textit{et. al.}~\cite{long_geometry_2021}. This force is
\begin{equation} \label{eq:two_disclinations_PK}
    {\bf f}_\text{PK} = \frac{\pi k_1 k_2 K}{2s}{\bf e}_x,
\end{equation}
with the force exerted on $L_1$ by $L_2$ being equal in magnitude and opposite in sign. We conclude that when $k_1=k_2$---the lines have the same winding---the force is repulsive, and when $k_1=-k_2$---the lines have opposite winding---the force is attractive. Notably, it is also completely insensitive to the parameter $q$. For an achiral material this prediction is borne out by both simulation and experiment. This same framework can be applied much more generally~\cite{long_applications_2024}, and yields results that are also (at least qualitatively) in agreement with experimental observations for achiral nematics. 

We now turn to the case of parallel disclination lines in a cholesteric, with $q_0 > 0$. Before we do any analysis, we see what happens in a numerical simulation by minimising the free energy. To simulate disclinations we cannot use the Frank energy~\eqref{eq:frank_energy}, although we will continue to employ this for our analysis. Instead we minimise the Landau--deGennes energy of the Q-tensor $Q = {\bf n} \otimes {\bf n} - I/3$ with the initial ${\bf n}$ being the director with two disclinations of opposite winding $\pm 1/2$ separated along the $x$-axis, exactly Eq.~\eqref{eq:two_disclinations}. The energy to be minimised is then~\cite{degennes_physics_2013}
\begin{equation}
    \begin{aligned}
    E_\text{LdG} &= a_0\int \frac{1}{2}\text{tr}\,Q^2 - \frac{1}{3}\text{tr}\, Q^3 + \frac{1}{4}\text{tr}\, Q^4 \\
    & \ \ \ + \frac{K}{2}\int |\nabla Q|^2 + 2q_0 Q \cdot \nabla \times Q + q_0^2.
    \end{aligned}
    \label{eq:landau_energy}
\end{equation}
Minimising the distortion part of this energy corresponds to the same dynamics as Eq.~\eqref{eq:EL}, but with the director ${\bf n}$ replaced by the Q-tensor to allow for disclinations. 

We consider a box of height $h = 50$ in the $z$ direction and length $2h$ in the $x$ and $y$ directions. To minimise the potentially-confounding effects of other forces we impose no boundary conditions, making the director periodic along the $z$ direction. For an achiral system, with $q_0=0$, the disclinations move closer together along the $x$-axis regardless of the value of the phase winding $q$ and with a speed that increases as they get closer together (i.e. as the separation $2s$ gets smaller), exactly as the expression~\eqref{eq:two_disclinations_PK} for the Peach--Koehler force predicts. 

To simulate a strongly chiral system we consider $q_0=2\pi/h$ so that the box length $h$ is equal to the pitch. We initialise the defects so that the separation between them is also $h$. Let us first examine the case of a $\tau$-line, $q=0$. Instead of moving towards one another and annihilating, the $+1/2$ disclination always moves along the direction of its `tail' while the $-1/2$ disclination remains fixed. Thus, depending on the relative orientation they either move apart, Fig.~\ref{fig1}(a), or towards one another, Fig.~\ref{fig1}(b). We can even rotate the profile by 90 degrees, in which case the $+1/2$ disclination stills moves along the tail, which is now in a direction orthogonal to the axis between the two disclinations. This shows a curious similarity with the behaviour in an achiral contractile active nematic, where the active forces drive motion of $+1/2$ defects, but not $-1/2$ defects, along their tails. In the active system, the broken rotational symmetry of the $+1/2$ causes it to self-propel in this way, while the $-1/2$ has a three-fold rotational symmetry that results in a flow field that makes it stable~\cite{giomi_defect_2014}. As we shall argue below, similar symmetry considerations apply here. Of course, our system is passive and not active and so the dynamics does not continue indefinitely, and we eventually reach an equilibrium state in which the disclination positions are fixed. Notably, they do not annihilate as they do in the achiral system, even when the relative orientation is such that the motion of the motile defect is towards the other defect, as in Fig.~\ref{fig1}(b).

Another interesting behaviour occurs. In our simulations the $+1/2$ disclination rapidly converts into a $-1/2$ disclination via the emission of a $+1$-winding singular line which immediately `escapes into the third dimension' to give a meron tube~\cite{pollard_escape_2024, pollard_morse_2024}, the mechanism for generating double-twist regions that we identified in Section~\ref{sec:cholesterics}. The texture is initially tight, but becomes overtwisted through this process while still maintaining a consistent sense of handedness, ${\bf n}\cdot \nabla \times {\bf n} < 0$. 

\begin{figure*}[t]
\centering
\includegraphics[width=0.98\textwidth]{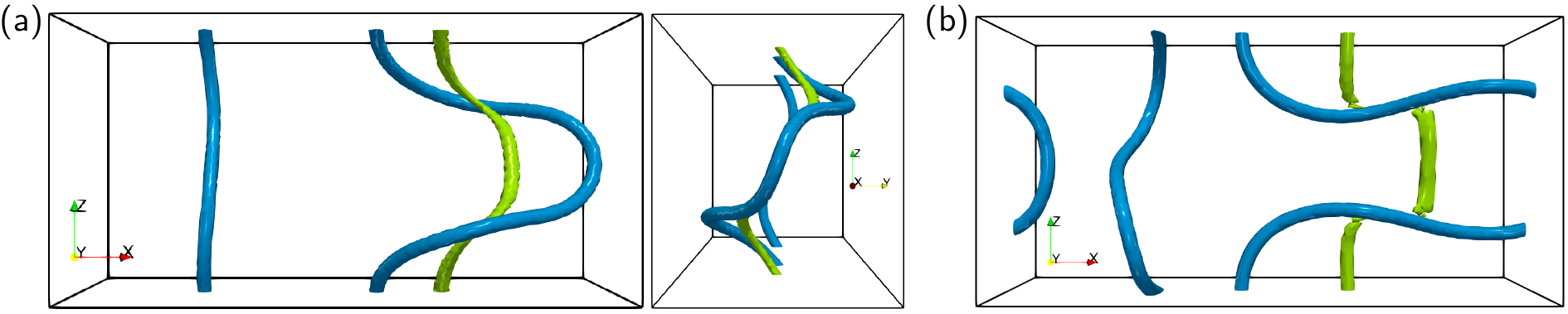}
\caption{Initially-straight $\chi$-lines in a cholesteric destabilise into helices. As with $\tau$-lines, defects with winding $-1/2$ (blue) are stable while those with winding $+1/2$ convert into winding $-1/2$ by the emission of a meron line (green), with the local structure of a double-twist cylinder. (a) The helix of the initially $+1/2$-winding line (right side of image) is wider than that of the initially $-1/2$-winding line (left side of image). The meron also has a helical shape. The helices are not perfect, but exhibit `kinks', as seen from the side-on view. (b) Running the simulation for longer results in the meron line breaking up into strands that connect parts of the $-1/2$ disclination. The simulation is shown for pitch $p$ equal to the height of the simulation box $h$.}
\label{fig2}
\end{figure*}

Now we simulate a $\chi$-line. We again use the director defined by Eq.~\eqref{eq:two_disclinations}, now with $q = 1$. We continue to use $q_0=2\pi/h$. The result of our simulation is shown in Fig.~\ref{fig2}. In this case we see a destabilisation of the $+1/2$ defect into a helix due to the motion of the defect along the `tail', whose direction now rotates as we move along the $z$-axis. As with the $\tau$-line, in this simulation the $+1/2$ singular line converts into a $-1/2$ winding singular line, leaving behind a $+1$ winding meron which sits at the core of the helix---this meron also has a slight helical winding to it, paralleling the disclination, which suggests the motion of the defect occurs slightly before the conversion process. The $-1/2$ defect also destabilises into a narrow helix, but the width of the helix does not grow as fast as that for the initially $+1/2$. Similar helical instabilities of $+1/2$ defect lines are also been observed in simulations of active cholesterics in thin films~\cite{wang_symmetry_2024}, without the conversion to the $-1/2$ defect occurring, again due to the self-propulsion effect of the $+1/2$ defect.

As time goes on the director field picks up a large $z$-component along the disclination that initially has $+1/2$ winding, but not along the disclination that initially has $-1/2$ winding. The shape is also not a perfect helix, but rather develops `kinks' that are clearly seen in Fig.~\ref{fig2}(a), with the helix becoming more flat in two places and more vertical in the middle. These particular geometric features have been observed in both the $\tau$- and $\lambda$-lines of chiral materials in various numerical and experimental settings~\cite{smalyukh_three-dimensional_2001, tai_three-dimensional_2019, wu_hopfions_2022, pieranski_cholesteric_2022, fukuda_exotic_2022}. They bear a striking resemblance to the `cusps' that appear in Legendrian curves---that is, curves whose tangent vector is everywhere orthogonal to the director---in contact topology~\cite{geiges_introduction_2008}. On the level of topology the disclinations in cholesterics themselves bear a close correspondence to Legendrian curves, with which they share a topological invariant called the Thurston--Bennequin number~\cite{geiges_introduction_2008, machon_contact_2017, pollard_contact_2023}. This half-integer is the number of $2\pi$ rotations of the director as we go around the disclination line, which defines the helical winding $q$ of a chiral $\chi$-line. It is invariant precisely because there are always Legendrian curves parallel to the disclination in arbitrarily small neighbourhood of the line with Thurston--Bennequin number $q$, and this cannot be changed by a homotopy preserving chirality~\cite{pollard_contact_2023}. The buckling of the $\chi$-line into a helix can then be seen as a buckling instability of the shape of cholesteric layers themselves, which is qualitatively similar to the Helfrich--Hurault instability~\cite{machon_contact_2017}. Eventually, the disclination crosses though the meron tube, resulting in a disclination with meron tethers, Fig.~\ref{fig2}(b). The local structure of these crossing points is of type $D_6^+$ in the terminology introduced in our previous work~\cite{pollard_morse_2024}. 

We have described this behaviour at fixed $q_0$, equivalently fixed pitch length, but it is natural to study the crossover between this behaviour and the behaviour of the achiral case $q_0=0$, where the disclination lines with opposite winding always come together and annihilate in the centre of the box in accordance with the predictions of the Peach--Koehler theory. To do this we vary the pitch length, so that $p = ah$ for $a \leq 1$ (i.e., the simulations shown in Fig.~\ref{fig1}, and Fig.~\ref{fig2} are at $a=1$). We find that the crossover occurs at $a \approx 0.2$: for values of $a \leq 0.2$ the $+1/2$ disclination does not change its winding and the disclinations come together and merge as they do in the achiral system.

The same behaviours observed at $p \sim h$ can be seen when we simulate a single disclination line in the box, with fixed boundary conditions, and thus the observed behaviour does not arise from interactions between disclination lines. We may suspect that an important role is played by the additional force ${\bf f}_\text{ch}$ of Eq.~\ref{eq:stress_force_chiral}, or rather the related torque ${\bf n} \times {\bf f}_\text{ch}$. 

We can confirm this by a computation of the Peach--Koehler force. Staring from the same director field as in the achiral case, we may compute its associated stress tensor~\eqref{eq:stress_chiral} and insert into the formula~\eqref{eq:peach-koehler}. Clearly, this will yield the same force on $L_2$ as in the achiral material, plus an extra term of the form
\begin{equation}
    -Kq_0\left( 2\pi k_1 {\bf e}_z  \cdot (I-{\bf n} {\bf n}) \right) \times {\bf e}_z.
\end{equation}
We see that, for the director field of the form ${\bf n} = \cos \phi \, {\bf e}_x + \sin \phi \, {\bf e}_y$ with $\phi$ as in Eq.~\eqref{eq:two_disclinations}, this extra term vanishes. 

A difficulty here is that our model director director field, which is harmonic, is an energy minimiser for an achiral material but not a chiral one. Finding an energy minimiser means solving the Euler--Lagrange equation~\eqref{eq:EL}, which is significantly more difficult than constructing a harmonic director field. Thus, it is quite reasonable that our calculation yields an incorrect prediction. We do not know how to write down a director field that exactly minimises the chiral energy. However, based on theory and a qualitative analysis of numerical simulations the director field 
\begin{equation} \label{eq:tau_line}
    {\bf n}^\tau_{1/2} = \cos(\theta/2){\bf m} - \sin (\theta/2) {\bf e}_z,
\end{equation}
with ${\bf m} = \cos(\theta/2) {\bf e}_x + \sin(\theta/2){\bf e}_y$, is reasonably close to a minimiser for the director field around a $+1/2$ $\tau$-line, where $\theta$ is the polar angle around the line~\cite{pollard_contact_2023}. The defect has tangent vector along the $z$ direction, and Burgers vector $\pi {\bf m}_\perp = \pi {\bf e}_z \times {\bf m}$. Let us treat this as the local structure for a $+1/2$ defect along $L_2$, alongside a $-1/2$ defect along $L_1$. We can calculate the Peach--Koehler force on the $-1/2$ defect due to the stress field generated by a $+1/2$ defect with local director field~\eqref{eq:tau_line}, and then exploit the fact that the force on the $+1/2$ must be equal and opposite. 

Because both the Burgers' vector and tangent to $L_1$ are aligned with ${\bf e}_z$, the only parts of the stress tensor that contribute to the Peach--Koehler force are $\sigma_{zx}, \sigma_{zy}$. We find that:
\begin{equation}
    \begin{aligned}
        \sigma_{zx} &= K([\cos^2(\theta/2)\cos 2\theta + \frac{1}{2}\sin^22\theta]\partial_x \theta \\
        & \ \ \ \ \  + q_0 \cos^2(\theta/2)\sin(\theta/2) ), \\
        \sigma_{zy} &= K([\cos^2(\theta/2)\cos 2\theta + \frac{1}{2}\sin^2 2\theta]\partial_y \theta \\
        & \ \ \  \ \ + q_0 \cos(\theta/2)\sin^2(\theta/2) ),
    \end{aligned}
\end{equation}
If we evaluate these along $L_1$ (i.e., at $x=-s, y=0, \theta = \pi$), we again find that the chiral terms disappear and the Peach--Koehler theory predicts an attractive force between the two disclinations along the $x$-axis. Thus, it is clear we need something else to explain the observed behaviour. 

We may also examine the behaviour of disclination lines in a more general setting. Simulating a quench, in which a material is rapidly cooled from the isotropic phase into the cholesteric phase, can be done by initialising the Q-tensor to be fully random inside a box. We impose planar anchoring on the top and bottom slices of the box, with a fixed director field ${\bf e}_x$ along the bottom and top slices. The material then nucleates a complex network of disclination lines, with no preference for a particular winding of the profile and no consistent sense of handedness. This complex network gradually coarsens away as the material minimises the energy~\eqref{eq:landau_energy}. At sufficiently high chirality, $q_0 \sim 2\pi/h$ for $h$ the box length, we observe similar behaviour to what is shown in Figs.~\ref{fig1} and~\ref{fig2}. Defects with $+1/2$ winding are rare, and the material prefers to convert them into $-1/2$ winding by the emission of meron lines. Over time, these disordered networks of disclinations and merons tend towards more symmetric structures, for example lattices of $-1/2$ $\chi$-type disclinations with helical structure, connected by meron tubes (in a manner similar to what is shown in Fig.~\ref{fig2}(b)) that are required to meet topological constraints on the projection of the director into cross-sections. These structures have been observed previously in simulations~\cite{fukuda_quasi-two-dimensional_2011, kwok_cholesteric_2013} and are structurally similar to both meron lattices and to the blue phases~\cite{wright_crystalline_1989}. 

Plainly, our observed behaviour---$+1/2$ disclinations changing their winding via meron emission, rather than annihilating with $-1/2$ disclination lines---occurs quite generally, and also cannot be explained by the Peach--Koehler theory alone. In the next section we analyse this behaviour using an alternative approach.

\section{The Gray Stability Theorem and the Velocity of a Cholesteric Defect}
\label{sec:gray_stability}
We now give an analysis of the motion of cholesteric disclinations using a result from contact topology which allows us to directly calculate the velocity field in a chiral material, and also to do this globally and not just along defect lines. Direct calculations of defect velocity have been given in a general setting~\cite{schimming_kinematics_2023}, but the particular topological structure in a cholesteric vastly simplifies these calculations~\cite{machon_contact_2017}.  

\subsection{Evolution of Chiral Director Fields Without Defects}
Both fluids and elastic solids evolve by isotopy. This means the configuration of the material at time $t$ is related to its configuration at time $0$ by some time-parametrised diffeomorphism $\psi_t$, such that $\psi_0$ is the identity. Intuitively, we think of the actual molecules of the solid being moved to different places: the molecule initially at some position $p$ moves to the position $\psi_t(p)$ at time $t$. The time derivative ${\bf v}_t = \dot{\psi}_t \circ \psi_t^{-1}$ of this diffeomorphism gives the velocity of the material---in fluid dynamics, the diffeomorphism $\psi_t$ is called the Lagrangian flow, while ${\bf v}_t$ is the Eulerian flow. 

A liquid crystal behaves differently. In a passive nematic, the material `sticks' may rotate independently from the background space, inducing distortions without the actual `molecules' of the material moving from their initial positions, and hence the material evolution is via the weaker notion of homotopy. In a homotopy, we may imagine that each `stick' is pinned in place but may rotate freely, as opposed to an isotopy where the sticks may move but not rotate. Homotopy results in more complicated dynamics in general, however, the special topology of the cholesteric simplifies matters considerably. As long as the material is defect free and the constraint ${\bf n} \cdot \nabla \times {\bf n} \neq 0$ on the twist holds then, for any homotopy of the director that preserves this constraint, we can always find an isotopy that describes the same motion as the homotopy---that is, we can always think of the evolution of the director as being the result of the background molecules moving without any independent rotation, even if this is not physically what is occurring. This result, know as the Gray stability theorem in contact topology~\cite{gray_global_1959, geiges_introduction_2008, machon_contact_2017}, strongly constrains the motion of a chiral material. Furthermore, the theorem gives an explicit construction of the velocity field of the isotopy, as we now explain. 

For a nematic, an evolution via isotopy implies that there is a time-dependent diffeomorphism $\psi_t$ relating the director at time $t$, ${\bf n}_t$, to the initial director ${\bf n}_0$. This relationship is $\psi^*_t {\bf n}_t = {\bf n}_0$. The dual 1-form $\eta_t$ to the director field then satisfies the equation
\begin{equation}
    \psi^*_t \partial_t \eta_t = \lambda_t \eta_0,
\end{equation}
for some scale factor $\lambda_t$. By taking the time derivative, we find the velocity field ${\bf v}_t$ of $\psi_t$ satisfies~\cite{geiges_introduction_2008, machon_contact_2017}
\begin{equation} \label{eq:advection}
    \partial_t \eta + L_{\bf v} \eta = (\partial_t \log\lambda) \, \eta. 
\end{equation}
To determine ${\bf v}$ from Eq.\eqref{eq:advection} we assume ${\bf v}$ is orthogonal to ${\bf n}$ and also that ${\bf n} \cdot \nabla \times {\bf n} \neq 0$. Then we may solve explicitly, and we find the velocity field is given by, 
\begin{equation} \label{eq:flow_field1}
    {\bf v} = \frac{1}{{\bf n} \cdot \nabla \times {\bf n}} {\bf n} \times \partial_t {\bf n}. 
\end{equation}
See the textbook of Geiges~\cite{geiges_introduction_2008} or the paper of Machon~\cite{machon_contact_2017} for a full derivation. Importantly, we see that the flow field is invariant under ${\bf n} \mapsto -{\bf n}$, and hence its direction is unambiguous. As noted by Machon~\cite{machon_contact_2017}, this evolution also arises from an appropriate limit of the Ericksen--Leslie equations, and so can viewed as a kind of simplified hydrodynamics for a chiral director. It implies the conservation of the cholesteric layer structure, and can be used to explain the Helfrich--Hurault buckling instability in cholesterics~\cite{machon_contact_2017}. 

We should be careful to note that, while the flow field ${\bf v}$ can be defined for any cholesteric director, in general it will not correspond to the physical flow field. In particular, there is no density in our model for the liquid crystal, so it does not makes sense to impose an incompressibility constraint $\nabla \cdot {\bf v} = 0$, nor is it required as ${\bf v}$ is nonphysical. The divergence of the flow field does have meaning however: regions in which $\nabla \cdot {\bf v} > 0$ are growing, and those with $\nabla \cdot {\bf v} < 0$ are shrinking. 

The flow field~\eqref{eq:flow_field1} is valid for {\it any} evolution $\partial_t {\bf n}$, provided that the twist is nonvanishing at all times and no defects are present. Under dynamics corresponding to energy minimisation we know that the time evolution of a cholesteric director is given by the Euler--Lagrange equation~\eqref{eq:EL}. Putting~\eqref{eq:EL} into~\eqref{eq:flow_field1} then results in an exact global expression for the velocity ${\bf v}$ of the material's flow field:
\begin{equation} \label{eq:flow_field}
    {\bf v} = \frac{K}{{\bf n} \cdot \nabla \times {\bf n}} \left( {\bf n} \times \nabla^2{\bf n} - 2q_0 {\bf n} \times \nabla \times {\bf n}\right)
\end{equation}
For further geometric insight, we note the following relationship, 
\begin{equation}
    \nabla \times {\bf n} = ({\bf n} \cdot \nabla \times {\bf n}){\bf n} + {\bf n} \times {\bf b},
\end{equation}
where ${\bf b} = {\bf n} \cdot \nabla {\bf n}$ is the bend distortion. Thus, the contribution the curl term $-2q_0K \nabla \times {\bf n}$ in the Euler--Lagrange equation makes to the flow field \eqref{eq:flow_field} is aligned with ${\bf b}$. This gives an alternative formula for the flow field in terms of the bend distortion, 
\begin{equation} 
    {\bf v} = \frac{K}{{\bf n} \cdot \nabla \times {\bf n}} \left( {\bf n} \times \nabla^2{\bf n} + 2q_0 {\bf b}\right). 
\end{equation}
In many cases of interest the director will be described by a Laplacian eigendirection, $\nabla^2 {\bf n} = \lambda {\bf n}$. In this case, the flow field is determined entirely by the symmetry-breaking force that arises from chirality, 
\begin{equation}
    {\bf v} = -\frac{2Kq_0}{{\bf n} \cdot \nabla \times {\bf n}} {\bf n} \times {\bf f}_\text{ch}. 
\end{equation}

We can gain further insight into the behaviour of this flow field by examining a simple model. Consider the director field
\begin{equation} 
    {\bf n} = \cos f(z) \, {\bf e}_x + \sin f(z) \, {\bf e}_y,
\end{equation}
for $f$ an arbitrary function of $z$. The director rotates along the ${\bf e}_z$ direction. The standard cholesteric helix corresponds to a linear function $f(z) = qz$, which describes rotation of the director field at a constant rate $q$. For this more general director field we have 
\begin{equation}
    \begin{aligned}
        \nabla^2 {\bf n} &= \partial_z^2 f \, {\bf e}_z \times {\bf n} - (\partial_z f)^2{\bf n}, \\
        \nabla \times {\bf n} &= -\partial_z f \, {\bf n}.
    \end{aligned}
\end{equation}
This leads to the flow field,
\begin{equation}
    {\bf v} = K\frac{\partial_z^2 f}{\partial_z f}{\bf e}_z. 
\end{equation}
This vanishes for a constant rate of rotation (i.e. when $f$ is linear). When the rate of rotation is not constant the flow field acts to correct this by smoothing out gradients in $f$ along the rotation direction, the $z$-axis. Assuming the director field is constrained to be of this form and all time-dependence in the director is in the function $f$, the evolution equation for the director field under energy minimisation reduces to the heat equation $\partial_t f = K\partial^2_z f$, which very obviously has the effect of smoothing out gradients in the rate of rotation. 

\begin{figure}[t]
\centering
\includegraphics[width=0.4\textwidth]{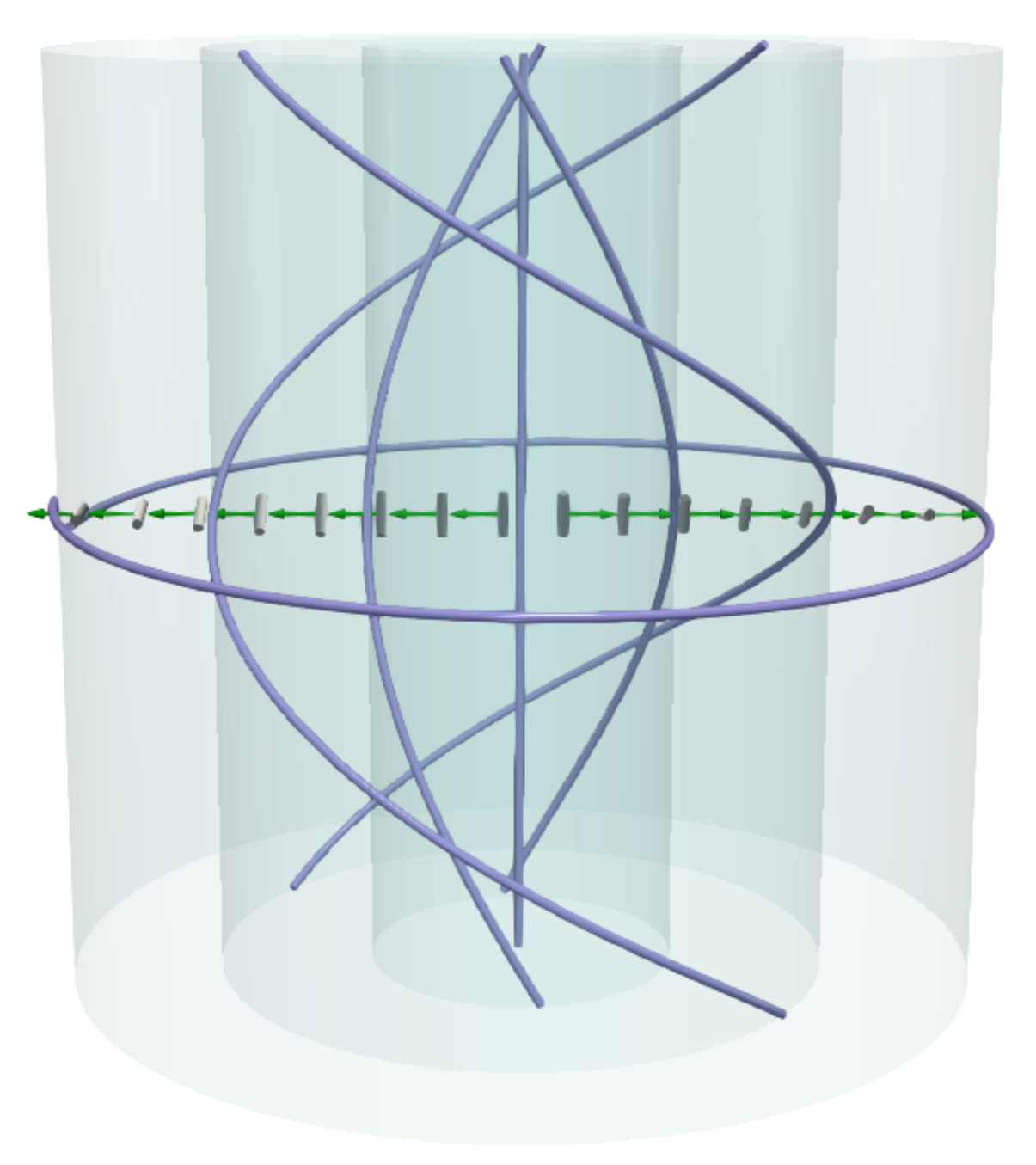}
\caption{Illustration of the Gray stability theorem for a meron in the form of a double-twist cylinder. Gray stability shows an equivalence between homotopies of the director field and isotopies, which can be viewed as moving the integral curves of the director field. In a meron, the integral curves of the director (purple curves) are helices that limit on a flat circles at the boundary of the meron and a straight line at its core. Energy minimisation leads to an expansion of the meron. This involves a homotopy of the director field (white), but Gray stability shows the result of this homotopy is the same as the flow of a diffeomorphism that translates the helices outwards along the velocity field in equation~\eqref{eq:flow_field}, shown in green. Equivalently, we think of the cylinders on which these integral curves sit (blue) also being translated outwards, leading to the growth of the domains which they bound.}
\label{fig3}
\end{figure}

We further illustrate the Gray stability theorem in Fig.~\ref{fig3}, which shows the flow field in a meron tube. The expansion of the meron core region is effected by a homotopy of the director field which rotates the `sticks', but by Gray stability this can be seen as the result of an isotopy which `moves' the helical integral curves of the director outwards.

\subsection{The Flow Field in the Presence of Defects}
This construction assumes that no defects are present. If there are point or line defects in the material, then the construction remains valid provided that the defects themselves only move by isotopy and do not undergo structural changes, such as a change in winding number. In general, an evolution that is a homotopy but not an isotopy is associated with one of the following structural changes to a material: (a) the creation or annihilation of defects, (b) a change in the structure of the defect set, for example a change in winding of a disclination line or the conversion of a point defect from a radial hedgehog to a hyperbolic hedgehog~\cite{pollard_morse_2024}, or else (c) the creation, annihilation, or topological reconfiguration of surfaces on which the twist vanishes. 

We can still use this theory to predict the motion of a defect line, computing the velocity field at any time except the instant in which structural changes (e.g., meron emission) occur. An advantage of this approach is that it does not just give us the motion of disclination lines but also point defects, as we discuss in Section~\ref{sec:hedgehog} below. We can also use it to track the motion of other topological/geometric structures associated with the cholesteric, such as merons, Legendrian curves, and the cholesteric layers~\cite{machon_contact_2017}.

It is important to note that the velocity field ${\bf v}$ of Eq.~\ref{eq:flow_field} will not generally be the velocity of the defects themselves. Rather, it describes the motion of the director field `sticks' via the relationship $\psi^*_t {\bf n}_t = {\bf n}_0$, and we may infer the motion of defects from this. 

We restrict ourselves to tight disclinations with a well-defined winding number along the entire length, and consider the two subclasses of tight disclination---the $\tau$- and $\chi$-lines---separately. The $\chi$-lines are the most straightforward. We consider the model director~\cite{pollard_contact_2023}, 
\begin{equation}  \label{eq:chiline}
    {\bf n}^\chi = \cos(k\theta + qz) {\bf e}_x + \sin(k\theta + qz) {\bf e}_y
\end{equation}
which has ${\bf n} \cdot \nabla \times {\bf n} = -q$ and describes a $\chi$-line with winding $k$. This director is harmonic, and hence the flow field is aligned with the bend direction. This leads to, 
\begin{equation}
    {\bf v}^\chi = -\frac{2kKq_0}{qr} \sin(qz + (k-1)\theta) {\bf n}^\chi_\perp, 
\end{equation}
where $r = \sqrt{x^2+y^2}$ and we have introduced ${\bf n}^\chi_\perp = {\bf e}_z \times {\bf n}^\chi$. 

\begin{figure*}[t]
\centering
\includegraphics[width=0.95\textwidth]{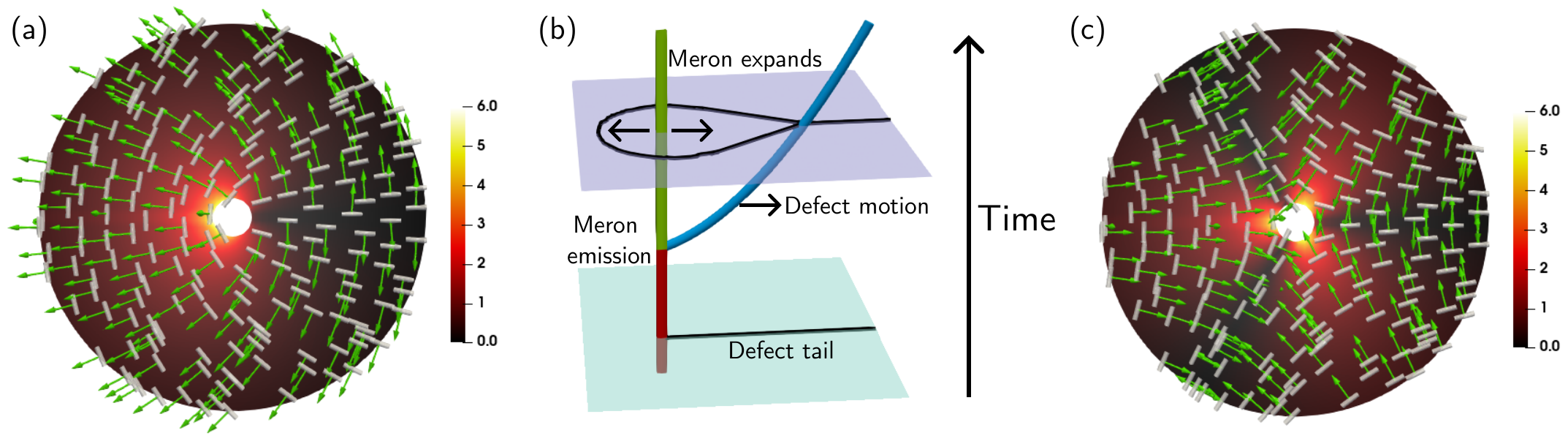}
\caption{(a) The material flow field generated by the director distortions around a $\chi$-line with winding $+1/2$. We show the flow field ${\bf v}$ (green) on the slice $z=0$ coloured by the magnitude $|{\bf v}|$, along with the director ${\bf n}$ (white). The flow depends on $z$ only through the orthogonal direction ${\bf n}_\perp$ to the director. (b) The action of this flow field causes the defect to convert into a $-1/2$ defect through the emission of a meron. We visualise this by examining the spatio-temporal change across a 2D slice in the material, with the third dimension in this panel corresponding to time. Initially---at a time corresponding to the pale blue slice---we see a $+1/2$ defect (red), which we visualise with the `tail' pointing right. As time goes on, the flow field causes the disclination to undergo the emission of a meron (green) and change its winding to $-1/2$ (blue). The resulting $-1/2$ defect is then pushed away by the expansion of the meron tube, as described in the text and indicated on the lilac slice. (c) The flow field around a $\chi$-line with winding $-1/2$. We can see visually that the symmetries of this flow field will not cause the defect to change its winding.}
\label{fig4}
\end{figure*}

These flow fields are shown in Fig.~\ref{fig4}, on the cross-section $z=0$---for different values of $z$ the picture is rotated as appropriate. In both cases, the velocity is not defined at the defect point itself. 

First consider the $+1/2$ disclination, Fig.~\ref{fig4}(a). Being aligned with ${\bf n}_\perp$, we see that the flow field vanishes at the defect point as well as along the `tail' of the defect, while simultaneously driving the director `sticks' away from the tail. One can see clearly how this flow pattern will drive a transition from $+1/2$-winding to $-1/2$-winding, by `opening up' the comet along the positive $x$-axis, resulting in meron emission. This process is visualised in Fig.~\ref{fig4}(b). We consider the defect profile on a single slice, the $z=0$ slice, and visualise the change in this profile over time (the third dimension in Fig.~\ref{fig4}(b)). We indicate the time 0 behaviour as a pale blue slice. Here, we see a single $+1/2$ defect (red). As time goes on, the flow field drives a transition into a pair consisting of a meron (double-twist cylinder, green) and a $-1/2$ defect (blue). In the terminology of our previous work~\cite{pollard_morse_2024} this is a $D_-^5$ (`parabolic umbilic') structural change in the defect set, which can also be observed in experiments~\cite{wu_topological_2024}. As we discuss presently, the flow field in the double-twist region surrounding the meron causes this region to expand, driving a motion of the now $-1/2$-winding defect---the behaviour along $z=0$ at this future time is shown on the lilac slice. The line across which the director is tangent to the $z=0$ is shown in black in Fig.~\ref{fig4}(b) at these two times. On the lilac slice, the black loop encloses the double-twist region, and the $-1/2$ defect is pinned to the boundary of this region. 

For a $\chi$-line the profile, and hence the direction of the `tail', rotates along the line, and hence we do indeed see the predicted destabilisation of an initially straight line into a helix. The handedness of the helix matches the handedness of the material---indeed, it changes handedness if we simultaneously change the sign of $q_0$ and $q$. 

By contrast, the flow pattern around the $-1/2$ defect, Fig.~\ref{fig4}(c), does not lead to motion. The flows drive the `sticks' inwards, towards the three `arms' of the $-1/2$ defect and the defect point itself, stabilising the structure. As we have remarked, this behaviour---$+1/2$ winding defects driving the motion, $-1/2$ defects remaining fixed---occurs due to the symmetry breaking of the $+1/2$ defect, the same reason as in an active nematic, albeit by a different mechanism. 

It is the meron (double-twist cylinder) that is born from this process that drives the motion of the $+1/2$ defect. The director field for a double-twist cylinder of radius $R$ is 
\begin{equation}
    {\bf n}_\text{DT} = \sin\left(\frac{\pi r}{2R}\right) {\bf e}_\theta - \cos\left(\frac{\pi r}{2R}\right) {\bf e}_z. 
\end{equation}
Note that the meron must be `escaped down', pointing along the negative $z$ axis at $r=0$, in order to be right-handed~\cite{pollard_escape_2024}, and for a left-handed version we would instead `escape up'. This changes the sign of various terms in the following computation, but these are cancelled out by a change in the sign of $q_0$ necessary for a left-handed material, and the conclusions remain the same.

We compute the curl, 
\begin{equation}
    \nabla \times {\bf n}_\text{DT} = - \frac{\pi}{2R}{\bf n}_\text{DT} + \frac{1}{r} \sin\left(\frac{\pi r}{2R}\right) {\bf e}_z.
\end{equation}
Since this director field is a Laplacian eigenvector, the flow field in the cylinder is determined entirely by the curl, 
\begin{equation}
    {\bf v}_\text{DT} = \frac{Kq_0R}{\pi r}\sin^2\left(\frac{\pi r}{2R}\right){\bf e}_r.
\end{equation}
Since the flow field is always directed radially outward we see it acts to drive material outwards, expanding the size of the region of double-twist. The disclination which emitted this meron is bound to the surface of this region, and therefore driven outwards (along the direction which was initially the `tail' of the $+1/2$ defect) by the expanding region of double-twist. Unlike an active nematic, the defect will not propel indefinitely as the system eventually reaches an equilibrium, corresponding to a desired radius for the emitted meron tube. We can estimate this equilibrium radius by calculating the energy, which was originally done by Wright \& Mermin in their analysis of the cholesteric blue phases~\cite{wright_crystalline_1989}. Making use of the cholesteric connection $D$ of Eq.~\eqref{eq:cholesteric_connection} allows us to compute the energy on a cross-section~\eqref{eq:cholesteric_energy}:
\begin{equation}
    \int_0^R | D{\bf n}_\text{DT}|^2 dr = \frac{8q_0^2R^2 + 4\pi q_0R + (2\pi + 4Rq_0)\text{Si}(\pi) - 4}{R},
\end{equation}
where $\text{Si}(x) = \int_0^x \sin t /t$. By taking the derivative with respect to $R$ and setting this equal to zero, we find this is minimised when, 
\begin{equation}
    8q_0^2R^2 = 2\pi\text{Si}(\pi) + \pi^2 -4.
\end{equation}
From this expression we find that the optimal value of the dimensionless quantity $q_0 R$ is approximately $1.479$. In terms of the pitch length $p = 2\pi/q_0$ we then have an optimal radius of $R \sim 0.235p$. This is consistent with the naive estimate that the radius of a double-twist cylinder is approximately a quarter of the cholesteric pitch length, and with what we seen in simulations. 

From this energetic argument we might naively imagine that a standard cholesteric helix will spontaneously nucleate merons. However, contact topology shows that a conversion from a tight cholesteric helix to an overtwisted state containing a double-twist cylinder requires either (a) the creation of defects, or (b) the nucleation of regions of reversed handedness. Moreover, since merons are themselves topological objects they can only be created in pairs, so the creation of double-twist cylinders also necessitates the nucleation of $-1$-winding merons in addition to regions of reversed handedness. This would also be required by a $-1/2$-winding disclination changing its winding to $+1/2$. 

A local model for a chiral $-1$-winding meron close to its core axis is
\begin{equation}
    {\bf n}_{-1} = \sin(\beta)(\cos \theta {\bf e}_x + \sin \theta {\bf e}_y) - \cos(\beta){\bf e}_z,
\end{equation}
with an approximate angle $\beta(r,\theta) = \tfrac{1}{2}r^2\sin(2\theta)$. Calculating $|D{\bf n}_\text{-1}|^2$ and restricting to $r=0$ gives the energy density along the central line,
\begin{equation}
    | D{\bf n}_\text{-1}|^2\big|_{r=0} = 2q_0^2,
\end{equation}
The energy density of the standard cholesteric helix vanishes: while a double-twist cylinder is energetically cheaper than a standard cholesteric helix close to its core, a $-1$-meron is energetically more costly. For this reason---as well as the flow field for a $-1/2$ defect---we do not expect a $-1/2$ disclination to change its profile by meron emission, and nor do we expect the spontaneous nucleation of double-twist regions in the standard helix, since these require the creation of $-1$ merons as well. 

The emission of a Neel meron (with a radial profile) from a $+1/2$ defect will also convert it to $-1/2$, but---in contrast to the double-twist cylinder---a Neel meron will be emitted from the `tail' of the $+1/2$ comet shape and cause the defect to propel along the `head'. We do not expect to observe this process in a cholesteric because Neel merons have higher energy than double-twist cylinders. However, we suggest that materials that have an energetic preference for splay distortions would exhibit similar phenomena to those described here for cholesterics, with $+1/2$ disclination lines changing their winding to $-1/2$ by the emission of Neel, rather than Bloch, merons.

For completeness we examine the $\tau$-lines as well. These can be described by the director field
\begin{equation}
    {\bf n}^\tau = \cos \phi \, {\bf m} + \sin \phi \, {\bf e}_z,
\end{equation}
where ${\bf m} = \cos(k\theta) {\bf e}_x + \sin(k\theta) {\bf e}_y$ is a planar director encoding a defect of winding $k = \pm 1/2$ and $\phi$ is a phase that is decreasing along the pitch axis ${\bf e}_z \times {\bf m}$~\cite{pollard_contact_2023}. This gives a good approximation to numerically and experimentally obtained structure. This yields a flow field for the $\tau$-lines which is similar to those shown in Fig.~\ref{fig4}. The behaviour is then qualitatively similar to the $\chi$-line, except we do not see the buckling into a helix because there is no rotation in the $z$-direction. 

Thus, it seems that this process of $+1/2$ defect lines changing their winding may be generic, and that it is the fundamental mechanism underlying the nucleation of double-twist regions within a cholesteric. Moreover, the Peach--Koehler theory cannot account for these changes. 

\begin{figure*}[t]
\centering
\includegraphics[width=0.98\textwidth]{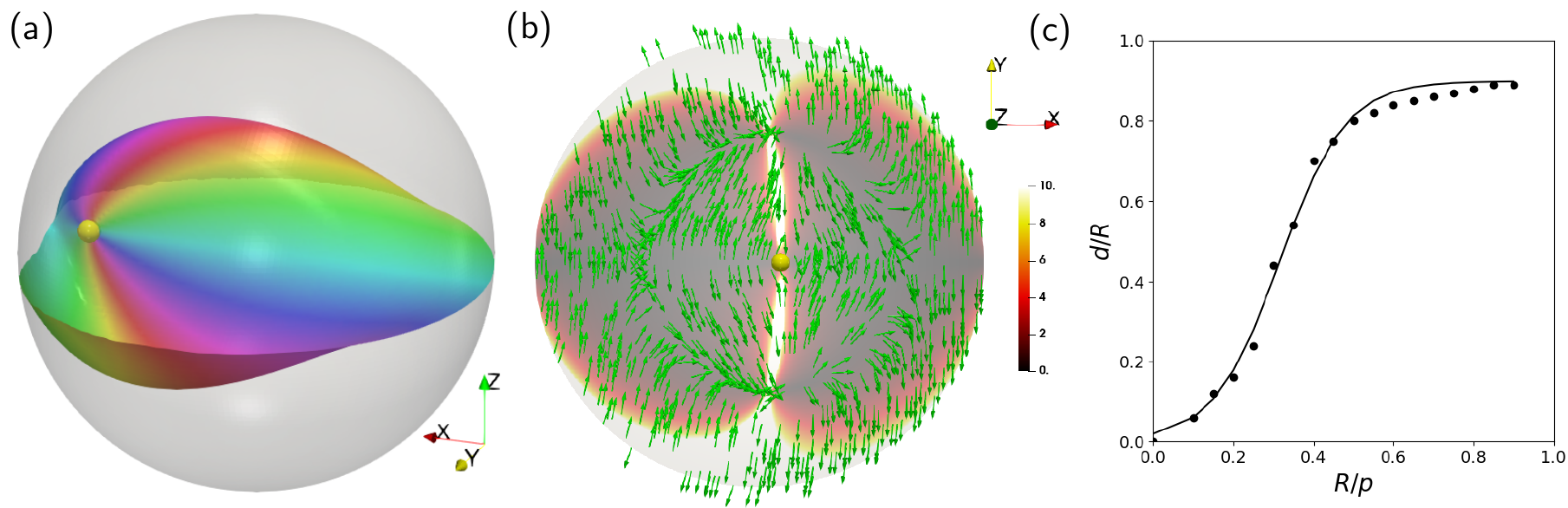}
\caption{(a) In a cholesteric, a hedgehog point defect initially at the centre of a spherical droplet with radial anchoring will displace from the centre of the droplet. We show the position of the point defect as a yellow orb, and we show a Pontryagin--Thom surface where the $z$ component of the director vanishes, coloured according to the angle between the $x$ and $y$ components. (b) This behaviour can be deduced from the projection into the $z=0$ surface of the flow field ${\bf v}$ (green), which predicts a flow of material away from the negative $x$-axis of the droplet, resulting in a motion of the point defect along the positive $x$-axis. The surface $z=0$, is coloured according to $|{\bf v}|$. (c) At a relatively low chirality (panel (a) shows $p = 0.5R$) this results in a equilibrium position of the defect that sits some distance $d > 0$ away from the centre of the droplet. We plot the relative displacement distance $d/R$ is a function of the ratio $R/p$ between the radius of the droplet and the pitch length (black points), which we fit to a sigmoid function, Eq.~\eqref{eq:fit} (solid black line). As the cholesteric pitch length grows the defect moves closer and closer to the boundary to minimise the size of the region of `wrong' handedness attached to it. For $R/p \geq 1$ the hedgehog reconfigures into a hyperbolic defect and moves back towards the centre of the droplet, as discussed in previous work~\cite{pollard_point_2019} (not illustrated). }
\label{fig5}
\end{figure*}

\section{The Motion of Point Defects}
\label{sec:hedgehog}
It is also interesting to apply this framework to the study of point defects in cholesterics. We consider a particular case of relevance to recent experiments: the behaviour of a radial point defect inside a spherical droplet~\cite{posnjak_hidden_2017, pollard_point_2019}. Since this situation involves no disclination lines, without loss of generality we may simulate this by minimising the Frank energy~\eqref{eq:frank_energy} rather than the Landau--deGennes energy~\eqref{eq:landau_energy}. Indeed, for the latter energy point defects are known to blow up into small rings, which is undesirable here.

We consider a spherical droplet of radius $R$ with normal anchoring boundary conditions. For an achiral system, the energy minimiser is a radial point defect at the centre of the droplet. Using arguments from contact topology, one can show that it is impossible for a director ${\bf n}$ containing a radial hedgehog point defect to be chiral, i.e. we cannot have ${\bf n} \cdot \nabla \times {\bf n} \neq 0$ in a neighbourhood of the defect, and hence hedgehogs always sit on a surface separating regions of reversed handedness~\cite{pollard_point_2019, pollard_escape_2024}. In experiments and simulations, hedgehogs in a spherical droplet of chiral nematic will displace from the centre of the droplet (their equilibrium position in an achiral nematic) and move towards the boundary in an attempt to minimise the region of disfavoured handedness. We show an example simulation with $R/p = 0.5$ in Fig.~\ref{fig5}(a), where $p = 2\pi/q_0$ is the pitch length. 

We can analyse this using the framework described in the previous section. The nematic hedgehog has ${\bf n} = {\bf e}_r$ in a spherical coordinate system $r, \theta, \phi$. Since the twist vanishes everywhere we cannot apply the framework described in the previous section to this director, so we instead consider a `twisted hedgehog' of the form,
\begin{equation} \label{eq:twisted_hedgehog}
    {\bf n}_\text{TH} = \frac{1}{r}\left( x{\bf e}_x - z{\bf e}_y + y{\bf e}_z \right)
\end{equation}
This director was introduced to approximate the local structure of surface-bound hedgehog defects in a cholesteric droplet, and it shows a good correspondence with experimentally observed textures~\cite{posnjak_hidden_2017, pollard_point_2019}. The twist is equal to $2x/r^2$, so the structure is left-handed (positive twist, energetically disfavoured) in the region $x > 0$ and right-handed (negative twist, energetically favoured) in the region $x < 0$. Informally, the twisted hedgehog director consists of a pair of double-twist cylinders of opposite directions of escape---equivalently opposite handedness~\cite{pollard_escape_2024}---meeting at the origin, which necessitates a defect due to the different topological classes of the escape up/down structures. 

While the twist vanishes along $x=0$ it is still possible for us to calculate the flow field within each half of the sphere that the twist is nonzero. We compute that $\nabla^2 {\bf n}_\text{TH} \sim {\bf n}_\text{TH}$, so again only the curl term will contribute to the flow field. We compute that 
\begin{equation}
    \begin{aligned}
    \nabla \times {\bf n}_\text{TH} &= \frac{1}{r^3}\bigg((2x^2+y^2+z^2){\bf e}_x + x(y-z){\bf e}_y \\
    & \ \ \ \ \ + x(y+z){\bf e}_z \bigg).
    \end{aligned}
\end{equation}
This leads us to 
\begin{equation}
    \begin{aligned}
    {\bf v}_\text{TH} &= \frac{Kq_0}{r^2}\bigg(w^2{\bf e}_x - (x(y-z)+ yw^2/x){\bf e}_y \\
    & \ \ \ \ \ - (x(y+z)+ zw^2/x){\bf e}_z \bigg).
    \end{aligned}
\end{equation}
where we have introduced $w = \sqrt{y^2+z^2}$. Close to the surface where $x=0$ we have $w \approx r$, and the ${\bf e}_x$-component of the velocity field (which, we observe, is well-defined at $x=0$ even if the ${\bf e}_y, {\bf e}_z$ components are not) is approximately $+Kq_0{\bf e}_x$. Thus, this surface should flow along the positive $x$-axis, and correspondingly the defect will also move away from the centre of the droplet along the positive $x$-axis with a speed proportional to $q_0$. This matches with the results of our numerical simulations. 

This calculation does not include the effects of a hard boundary condition, which is present in the experiments reported in Refs.~\cite{posnjak_hidden_2017, pollard_point_2019} as well as the simulation shown in Fig.~\ref{fig5}(a). This could be included as an additional force by introducing a term proportional to $(1-{\bf n}\cdot {\bf e}_r)^2$ into the energy, which penalises deviation from the radial anchoring boundary condition. We calculate the flow field numerically for such a texture, with the point defect displaced slightly along the positive $x$-axis, Fig.~\ref{fig5}(b). As our calculation suggests, the flow field drives a motion of the defect along the positive $x$-axis. The Pontryagin--Thom surface with $n_z=0$ develops a characteristic `clamshell' shape, seen in Fig.~\ref{fig5}(a). This too can be predicted from the flow field, which drives material `up' away from the $z=0$ plane on one side of the droplet, and `down' on the other.

This describes the instantaneous motion of the defect, but not its long term behaviour. The boundary repulsion effect results in an equilibrium position for the defect which is dependent on the inverse pitch length $q_0$. By performing simulations starting from the initial condition~\eqref{eq:twisted_hedgehog} with the defect displaced slightly from the origin, we determine the equilibrium normalised displacement $d/R$ of the hedgehog from the centre of the droplet as a function of the ratio $R/p$, for fixed $K = 0.1$ and a hard boundary condition, corresponding to an effectively infinite anchoring energy \footnote[4]{For an effectively infinite anchoring strength this distance will not depend on $K$, however for finite anchoring strength it will depend on the ratio between $K$ and the elastic constant associated with the anchoring term in the free energy.}. The numerically-calculated values are shown as circles in Fig.~\ref{fig5}(c). We fit the displacement $d/R$ against a sigmoid function,
\begin{equation} \label{eq:fit}
    \frac{d}{R} = \frac{0.9}{1+\exp(-a(x-b))}.
\end{equation}
We find values $a = 12.111, b = 0.316$. This curve is plotted in Fig.~\ref{fig5}(c). 

At large enough chirality, $p \sim R$, the hedgehog initially displaces towards the boundary, but then eventually moves back towards the centre of the droplet as it transforms from a radial to a hyperbolic structure which can be locally chiral~\cite{pollard_point_2019}. Various aspects of this process have been described previously~\cite{pollard_point_2019, paparini_spiralling_2023, ciuchi_inversion_2024, pollard_morse_2024}.

\section{Discussion}
\label{sec:discussion}
The energetic preference for maintaining a uniform sense of handedness imposes strong constraints on the textures formed in a chiral material, and in particular the structure of the defects~\cite{beller_geometry_2014, pollard_point_2019, pollard_contact_2023, pollard_escape_2024}. It also imposes strong constraints on dynamics~\cite{machon_contact_2017}, a fact we have exploited here to study the kinematics of defects in a passive cholesteric liquid crystal. We have shown via a combination of simulations and analytical calculations that the $\tau$- and $\chi$-line disclinations of a cholesteric exhibit qualitatively different dynamical behaviour from one another and also from the achiral nematic case. Notably, defects of opposite winding do not attract as they do in an achiral material, but rather $+1/2$ defects change their winding to $-1/2$ by emission of a meron, and the resulting disclination propels along the `tail' of the original $+1/2$ `comet'. This latter behaviour results in initially-straight $\chi$-lines destabilising into helices. Following Machon~\cite{machon_contact_2017}, we have performed an analysis of disclination motion in cholesterics using the Gray stability theorem from contact topology to calculate the material velocity field, which we have used to explain this phenomenon: the asymmetry of $+1/2$ defects drives an `unfolding' of the comet shape, resulting in change in winding and corresponding meron emission.

We have argued that the Peach--Koehler force does not explain this behaviour, as the additional terms that arise from chirality do not result in different predictions of the motion of the disclination lines. It is not clear to us whether this can be remedied, either by the consideration of additional forces, or perhaps a modification to the definition of the Burgers vector along the lines suggested in our previous work~\cite{pollard_morse_2024}. In general, we believe that the notion of meron-mediated interactions between disclination lines \cite{pollard_morse_2024} challenges the extent to which the Peach--Koehler formalism can describe the full behaviour of nematic liquid crystals.

It is well known that the Peach--Koehler force in both a crystal and a liquid crystal is analogous to the Lorentz force in electromagnetism~\cite{peach_forces_1950, degennes_physics_2013}. Our work therefore hints at the potential for further analogies between liquid crystals and well-studied phenomena in field theory and particle physics. We have described an interaction between disclinations which is mediated by merons and cannot be explained by the Peach--Koehler force---this sounds tantalisingly similar to the weak interaction between fermions, which does not arise from electromagnetism and is mediated by bosons. It is natural to ask whether this qualitative comparison may be made formal, and whether this may facilitate a deeper understanding of the phenomena we have described in this work. It may be possible to derive such a correspondence by considering field theories of liquid crystals. For example, the free energy of a uniaxial liquid crystal can be written as an $SO(2)$ gauge theory with an interaction term involving the Peach--Koehler force, at least when the material is achiral~\cite{kawasaki_gauge_1985}. Further, an analogy can be made between smectics/cholesterics and the Ginzburg--Landau theory of superconductors, and this analogy has been used to understand the twist-grain boundary phase~\cite{lubensky_tgb_1995}.

We have also applied our analysis based on the Gray stability theorem to a hedgehog point defect in a spherical droplet, a scenario where the assumption that the director is everywhere chiral does not hold. Even without this assumption, the calculation of the flow field from Gray stability is still possible, and it still sheds insight on the behaviour of the defect, which drifts away from the centre of the droplet and towards the boundary~\cite{posnjak_hidden_2017, pollard_point_2019}. Our calculations are based on the energy~\eqref{eq:frank_energy}, which is also the energy of a chiral ferromagnet with the Dzyaloshinskii-Moriya interaction. Our results can therefore be applied to under stand the motion of merons, Skyrmions and point defects (Bloch points) in these materials as well. 

The example of the hedgehog defect suggests that an analysis using the Gray stability theorem could potentially be extended far beyond its original application to defect-free chiral structures by relaxing the requirement that the twist be nonzero. More generally, the distinction between motion of a nematic which is a homotopy that integrates to an isotopy and a motion which is merely a homotopy is both subtle and interesting~\cite{pollard_morse_2024}, with structural and dynamical consequences that are not well understood. Violation of the integrability condition occurs due to the creation/annihilation of topologically-protected features (defects, solitons) and potentially also due to the creation/annihilation of non-topological features such as twist, splay, and bend walls. While bend and splay walls are frequently discussed in the context of active nematics due to their relationship with the active force, especially in two-dimensional systems, to our knowledge the dynamics of twist walls have not been examined in any detail, nor have two-dimensional bend and splay walls been analysed in detail in three-dimensional systems. The Gray stability theorem may help to study the motion of these geometric features in both passive and active materials, and we welcome further work in the area.

\section*{Acknowledgements}
JP and RGM acknowledge funding from the EMBL Australia program. RGM acknowledges funding from the Australian Research Council Centre of Excellence for Mathematical Analysis of Cellular Systems (CE230100001).

\end{document}